\documentclass[%
 aip,
 jmp,%
 amsmath,amssymb,
 reprint,%
 nofootinbib,%
 floatfix, %
 ]{revtex4-1}

\usepackage{graphicx}
\usepackage{caption}
\usepackage{subcaption}
\usepackage{dcolumn}
\usepackage{bm}
\usepackage[dvipsnames]{xcolor}
\usepackage{amssymb}
\usepackage{verbatim}
\usepackage{appendix}
\usepackage{listings}
\usepackage{units}
\usepackage{fancyhdr}
\usepackage{csquotes}
\usepackage{hyperref}

\usepackage{etoolbox}
\makeatletter
\patchcmd{\hyper@makecurrent}{%
    \ifx\Hy@param\Hy@chapterstring
        \let\Hy@param\Hy@chapapp
    \fi
}{%
    \iftoggle{inappendix}{
        \@checkappendixparam{chapter}%
        \@checkappendixparam{section}%
        \@checkappendixparam{subsection}%
        \@checkappendixparam{subsubsection}%
        \@checkappendixparam{paragraph}%
        \@checkappendixparam{subparagraph}%
    }{}%
}{}{\errmessage{failed to patch}}

\newcommand*{\@checkappendixparam}[1]{%
    \def\@checkappendixparamtmp{#1}%
    \ifx\Hy@param\@checkappendixparamtmp
        \let\Hy@param\Hy@appendixstring
    \fi
}
\makeatletter

\newtoggle{inappendix}
\togglefalse{inappendix}

\apptocmd{\appendix}{\toggletrue{inappendix}}{}{\errmessage{failed to patch}}
\apptocmd{\subappendices}{\toggletrue{inappendix}}{}{\errmessage{failed to patch}}

\newcommand*\obar[2][0.75]{
    \sbox{\myboxA}{$\m@th#2$}%
    \setbox\myboxB\null
    \ht\myboxB=\ht\myboxA%
    \dp\myboxB=\dp\myboxA%
    \wd\myboxB=#1\wd\myboxA
    \sbox\myboxB{$\m@th\overline{\copy\myboxB}$}
    \setlength\mylenA{\the\wd\myboxA}
    \addtolength\mylenA{-\the\wd\myboxB}%
    \ifdim\wd\myboxB<\wd\myboxA%
       \rlap{\hskip 0.5\mylenA\usebox\myboxB}{\usebox\myboxA}%
    \else
        \hskip -0.5\mylenA\rlap{\usebox\myboxA}{\hskip 0.5\mylenA\usebox\myboxB}%
    \fi}
\makeatother

\renewcommand{\vec}[1]{\boldsymbol{#1}}

\newcommand{\lang}{\left\langle}
\newcommand{\rang}{\right\rangle}

\renewcommand{\d}{\ensuremath{\mathrm{d}}}

\newcommand{\remove}[1]{{}}
\newcommand*{\nouncite}[1]{Ref.~\citenum{#1}}
\newcommand*{\nouncites}[1]{Refs.~\citenum{#1}}

\mathchardef\mhyphen="2D

\newcommand{\appref}[1]{\hyperref[#1]{Appendix~\ref{#1}}}

\usepackage{morefloats}
\begin{document}

\preprint{AIP/123-QED}

\title{Optimization of flux-surface density variation in stellarator plasmas with respect to the transport of collisional impurities}

\author{S.\ Buller}
\email{bstefan@chalmers.se}
\affiliation{Department of Physics, Chalmers University of Technology,
  SE-41296 G\"{o}teborg, Sweden}
\author{H.M.\ Smith}
\affiliation{Max-Planck-Institut f\"{u}r Plasmaphysik, 17491 Greifswald, Germany}
\author{A.\ Moll\'en}
\affiliation{Max-Planck-Institut f\"{u}r Plasmaphysik, 17491 Greifswald, Germany}
\author{S.L.\ Newton}
\affiliation{CCFE, Culham Centre for Fusion Energy, Abingdon, Oxon OX14 3DB, UK}
\affiliation{Department of Physics, Chalmers University of Technology,
  SE-41296 G\"{o}teborg, Sweden}
\author{I.\ Pusztai}
\affiliation{Department of Physics, Chalmers University of Technology,
  SE-41296 G\"{o}teborg, Sweden}

\date{\today}
\begin{abstract}
  Avoiding impurity accumulation is a requirement for steady-state stellarator operation. The accumulation of impurities can be heavily affected by variations in their density on the flux-surface. Using recently derived semi-analytic expressions for the transport of a collisional impurity species with high-$Z$ and flux-surface density-variation in the presence of a low-collisionality bulk ion species,
  we numerically optimize the impurity density-variation on the flux-surface to minimize the radial peaking factor of the impurities.
  These optimized density-variations can reduce the core impurity density by $0.75^Z$ (with $Z$ the impurity charge number) in the Large Helical Device case considered here, and by $0.89^Z$ in a Wendelstein 7-X standard configuration case.
  On the other hand, when the same procedure is used to find density-variations that maximize the peaking factor, it is notably increased compared to the case with no density-variation. This highlights the potential importance of measuring and controlling these variations in experiments. 
\end{abstract}

\maketitle

\section{Introduction}
\label{sec:intro}
Highly-charged impurities can enter fusion plasmas from the walls of the vessel. In any magnetic-confinement fusion-reactor, these impurities must be prevented from accumulating in the core of the plasma, where they radiate strongly and lead to unsustainable energy losses. 

Impurity accumulation is one of the major challenges in stellarators \cite{sudo2016}, where it is often observed when the radial electric field points inwards\cite{sudo2016,giannone2000,nakamura2002,burhenn2009}, which occurs when the electron and ion temperatures are comparable\cite{yokoyama2007}, and is a preferred regime for future reactors\cite{sagara20101336}. 

Such accumulation is also predicted theoretically, based on collisional transport calculations with simplified pitch-angle scattering collision operators. However, such simplified operators are not appropriate for treating highly-charged impurities with high collisionality\cite{helanderPRL2017}. When an appropriate collision operator is used, the radial transport of the heavy impurities becomes insensitive to the radial electric field\cite{braun2010a,helanderPRL2017,newton2017}. Additionally, if the bulk ions are in a low-collisionality regime, impurity accumulation can be avoided by having large temperature gradients\cite{helanderPRL2017,newton2017} -- an effect known as \emph{temperature screening}.

However, these results do not account for the tendency of highly-charged impurities to develop density-variations on the flux-surface\cite{helanderBifurcated1998,fulop1999,angioni2014}. Such variations occur when any plasma-species deviates from a flux-surface Maxwell-Boltzmann distribution, and are strongest for highly-charged species. When these variations are accounted for, the radial electric field again affects the transport of heavy-impurities\cite{calvo2018nf,buller2018jpp}. In most scenarios, this leads to an accumulation of impurities\cite{buller2018jpp,calvo2018nf}, but there exist cases where the radial electric field is mildly beneficial:
In \nouncite{calvo2018nf}, where the density variation is due to electrostatic potential-variation from trapped-particles\cite{calvo2018b} -- calculated with the \textsc{Euterpe} code -- there exists a narrow range of inward radial-electric fields that lead to an outward impurity flux for one of the charge states ($Z=24$, the lowest charge state investigated). 
In \nouncite{buller2018jpp}, where a model of a localized flux-surface variation of impurities was considered, it was found that an inward radial-electric field can lead to weak outward transport if the amplitude of the localized variation is small.
Furthermore, the outward transport of carbon impurities can be enhanced by electrostatic potential-variations caused by fast, perpendicularly injected, neutral-beam particles\cite{yamaguchiIAEA2018}. 

Thus, it is in some cases possible for flux-surface density variations to reduce or prevent impurity accumulation.
The aim of the present work is to theoretically explore what beneficial impurity density variations look like, and by how much they can reduce impurity accumulation.
We do this by optimizing the radial transport of impurities with respect to the impurity density variation, based on semi-analytical expressions for the impurity transport coefficients from \nouncite{buller2018jpp}. In particular, we look at variations that lead to the least (or most) impurity accumulation. These expressions account for the classical and neoclassical transport; recent studies indicate that turbulent transport may be significant for the impurities\cite{bgeiger2019}.
However, for highly charged impurities, even if the diffusive part of the transport is dominated by turbulence (as found in \nouncite{bgeiger2019}), the convective part of the transport (i.e. the one independent of the impurity density gradient) -- that determines the sign of the peaking factor -- can still be dominated by collisional transport, as it increases with $Z$.   
We do not include turbulent transport in this work, but note that it is also affected by flux surface variations\cite{fulop2011,mollen2012}.  

The remainder of this paper is organized as follows: in the next section, we present the mathematical formulation of the problem and analytic expressions for the impurity transport coefficients, radial flux and peaking factor.
In \autoref{sec:opt}, we explain how the impurity density variation is represented and optimized. In \autoref{sec:res}, we present the results of such optimization applied to model cases based on Wendelstein 7-X (W7-X) and Large Helical Device (LHD) profiles and magnetic geometries. These initial explorations indicate that tuning the impurity density variations could be useful in reducing core impurity accumulation. Further work is however needed to assess how the presence of turbulence would affect the results.

\section{Mathematical formulation}
\label{sec:mf}
In this section, we describe the radial transport of a highly-charged collisional impurity species (labeled ``$z$'') due to collisions with a bulk-ion species (labeled ``$i$''). Specifically, we consider a \emph{mixed-collisionality} regime\cite{helanderPRL2017} where the impurities are collisional and the bulk ions are in the $1/\nu$ regime. The impurities are not restricted to the trace limit, but are assumed to be highly charged, with charge $Ze$, where $Z \gg 1$ and $e$ is the proton charge.
We here summarize results from \nouncite{buller2018jpp}, which is a generalization of \nouncites{helanderBifurcated1998,fulop1999,angioni2014} from tokamaks to stellarators, and a generalization of \nouncites{helanderPRL2017,newton2017} for impurities varying on the flux surface.

The parallel momentum equation of a collisional impurity species is
\begin{equation}
T \nabla_\| n_z  =  -Z e n_z \nabla_\| \Phi + R_{z\|}, \label{eq:parallel}
\end{equation}
where $T$ is the impurity temperature, which is equilibrated with the main ion temperature $T_i = T$; $\nabla_\|$ denotes the gradient projected onto the magnetic field direction; $n_z$ is the impurity density, which varies along the flux-surface; $\Phi$ the electrostatic potential; and $R_{z\|}$ the parallel friction force. 
In the mixed-collisionality regime, the friction force is smaller than the other terms in this equation, which results in
\begin{equation}
n_z  =  N_z(r_N) \exp{\left(-\frac{Z e \tilde{\Phi}}{T}\right)}, \label{eq:boltzmann}
\end{equation}
where $r_N$ is a flux-surface label (in this paper, $r_N = \sqrt{\psi_\text{t}/\psi_{\text{t},\text{LCFS}}}$, where $\psi_t$ is the toroidal flux and $\psi_{\text{t},\text{LCFS}}$ the toroidal flux at the last-closed flux-surface); $N_z$ is a flux-function known as the \emph{pseudo-density}; and $\tilde{\Phi}$ is the deviation of $\Phi$ from its flux-surface average. The form of the density in \eqref{eq:boltzmann} is known as the \emph{Boltzmann-response} to the electrostatic potential, and is the equilibrium density in confined plasmas. If all species in the plasma obeyed \eqref{eq:boltzmann}, quasi-neutrality would force $\Phi$ to be a flux-function, and all densities would be constant along the flux-surface. However, if any species deviate from \eqref{eq:boltzmann}, $\Phi$ will vary on the flux-surface, which can cause species with high $Z$ to develop significant density-variations on the flux-surface.

Assuming that \eqref{eq:boltzmann} holds for the impurities -- regardless of the mechanism that causes the $\Phi$ variation -- the collisional radial transport of the impurities in the mixed-collisionality regime can be written as
\begin{equation}
\begin{aligned}
    \frac{\lang \vec{\Gamma}_z \cdot \nabla r_\text{N} \rang}{\lang n_z \rang} =& - D_{N_z}[n_z] \frac{1}{Z}\frac{\d \ln N_z}{\d r_\text{N}} + D_{\Phi}[n_z] \frac{e}{T} \frac{\d \langle\Phi\rangle}{\d r_\text{N}} \\&  - D_{n_i}[n_z] \frac{\d \ln n_i}{\d r_\text{N}} - D_{T_i}[n_z] \frac{\d \ln T}{\d r_\text{N}},
  \end{aligned} \label{eq:Gamma}
\end{equation}
where $\lang \cdot \rang$ denotes the flux-surface average.
In \eqref{eq:Gamma} the transport coefficients $D_X[n_z]$ depend on the impurity-variation on the flux-surface, as indicated by the square-brackets. Specifically,
\begin{widetext}
\begin{align}
  \frac{D_{N_z}}{D} = &
                         \lang n_z w^2 B^2 \rang  - \lang n_z wB^2 \rang\frac{\lang w B^2 \rang}{\lang B^2 \rang}  + \frac{\frac{\lang n_z w B^2 \rang}{\lang B^2\rang} \lang \frac{B^2}{n_z}\rang-\lang wB^2 \rang }{\lang \frac{B^2}{n_z} (1 - c_4 \alpha)\rang}  \lang (1 - c_4 \alpha)wB^2\rang + \lang \frac{n_z | \nabla r_N|^2}{B^2}\rang,\label{eq:DNz}\\
  \frac{D_{n_i}}{D} = &
                          -\lang n_z w u B^2\rang + \lang n_z w B^2\rang \frac{\lang u B^2 \rang}{\lang B^2 \rang}    
    -   \frac{\frac{\lang n_z w B^2 \rang}{\lang B^2\rang} \lang \frac{B^2}{n_z}\rang-\lang wB^2 \rang }{\lang \frac{B^2}{n_z} (1 - c_4 \alpha)\rang} \left(c_2 + \lang uB^2 \rang [c_1 +1]\right) 
    -\lang \frac{n_z | \nabla r_N|^2}{B^2}\rang,
   \label{eq:Dni}
\\
  \frac{D_{T_i}}{D} = &
\frac{1}{2} \lang n_z w u B^2\rang - \frac{1}{2}\lang n_z w B^2\rang \frac{\lang u B^2 \rang}{\lang B^2 \rang}                           
                                -\frac{\frac{\lang n_z w B^2 \rang}{\lang B^2\rang} \lang \frac{B^2}{n_z}\rang-\lang wB^2 \rang }{\lang \frac{B^2}{n_z} (1 - c_4 \alpha)\rang}
    \left(c_3 - \frac{3}{2} c_2 - \lang uB^2 \rang \left[c_1(\eta - 1) + \frac{1}{2}\right]\right)  \label{eq:DTi}
    \\&
        + \frac{1}{2}\lang \frac{n_z | \nabla r_N|^2}{B^2}\rang,\nonumber  \\
     D_{\Phi} = &-D_{n_i} - D_{N_z}, \label{eq:DPhi}
\end{align}
\end{widetext}
where $D=\frac{m_i n_i T_i }{Ze^2\lang n_z \rang n_z \tau_{iz}}$; $\alpha = Z^2 n_z/n_i$; $\eta\approx 1.17$; $c_1$ to $c_4$ are flux-surface constants that depend on the magnetic geometry and $n_z$, defined in the Appendices A--D of \nouncite{buller2018jpp}; and
\begin{align}
  \vec{B} \cdot \nabla u &= - \vec{B} \times \nabla \psi \cdot \nabla (B^{-2}), \label{eq:u}\\
  \vec{B} \cdot \nabla (n_z w) &= - \vec{B} \times \nabla \psi \cdot \nabla (n_z B^{-2}). \label{eq:w}
\end{align}
The functions $u$, $w$ are evaluated numerically from the magnetic geometry and $n_z$; the flux-surface constants $c_1$ to $c_4$ are integrated numerically. 

From \eqref{eq:Gamma}, we obtain the ($Z$-normalized) peaking factor of $N_z$ by solving for the $-\frac{1}{Z}\frac{\d \ln N_z}{\d r_\text{N}}$ that gives $\langle \vec{\Gamma}_z \cdot \nabla r_\text{N} \rangle =0$. Denoting this peaking factor by $\mathcal{P}$, we obtain
\begin{equation}
  \begin{aligned}
    \mathcal{P} =& -\frac{D_{\Phi}}{D_{N_z}} \frac{e}{T} \frac{\d \langle\Phi\rangle}{\d r_\text{N}} +  \frac{D_{n_i}}{D_{N_z}} \frac{\d \ln n_i}{\d r_\text{N}} + \frac{D_{T_i}}{D_{N_z}} \frac{\d \ln T}{\d r_\text{N}} \\
    =&\left(1+\frac{D_{n_i}}{D_{N_z}}\right) \frac{e}{T} \frac{\d \langle\Phi\rangle}{\d r_\text{N}} +  \frac{D_{n_i}}{D_{N_z}} \frac{\d \ln n_i}{\d r_\text{N}} + \frac{D_{T_i}}{D_{N_z}} \frac{\d \ln T}{\d r_\text{N}}.
  \end{aligned} \label{eq:P}
\end{equation}
The above definition of the peaking factor makes the problem independent of $Z$, but differs from the conventional definition of the peaking factor by a factor $1/Z$. As a result, the peaking of the actual $N_z$ profile for a given value of $\mathcal{P}$ increases with $Z$. Thus, even modest reductions in $\mathcal{P}$ can lead to large reductions in the core impurity content for large $Z$.

To find the $n_z$ that minimizes impurity accumulation, we can minimize $\mathcal{P}$ which, according to \eqref{eq:P}, is equivalent to minimizing a weighted sum of $D_{n_i}/D_{N_z}$ and $D_{T_i}/D_{N_z}$. 
This picture can be further simplified, as we often find that $D_{T_i} \approx -0.5 D_{n_i}$ to within a few percent\cite{buller2018jpp,bullerIAEA2018}. As a reference, we note that $D_{T_i} = -\frac{1}{2} D_{n_i}$ is exact for $n_z$ that are constant on the flux-surface, and that only the third terms in \eqref{eq:Dni} and \eqref{eq:DTi} cause deviation from this exact relation. 
 
With this approximation,
\begin{equation}
  \begin{aligned}
    \mathcal{P} 
    \approx &\frac{e}{T} \frac{\d \langle\Phi\rangle}{\d r_\text{N}}+\frac{D_{n_i}}{D_{N_z}} X,
  \end{aligned} \label{eq:approxP}
\end{equation}
where we have defined $X$ to be the combined thermodynamic force 
\begin{equation}
X \equiv \left(\frac{e}{T} \frac{\d \langle\Phi\rangle}{\d r_\text{N}}  +  \frac{\d \ln n_i}{\d r_\text{N}} -\frac{1}{2} \frac{\d \ln T}{\d r_\text{N}}\right). \label{eq:X}
\end{equation}
Thus, minimizing the peaking factor becomes equivalent to either maximizing or minimizing $D_{n_i}/D_{N_z}$, depending on the sign of $X$. When $X$ is far from zero, we hence expect the $n_z$ that optimizes the peaking factor to be insensitive to the radial gradients. This is potentially useful, as it would be cumbersome to calculate an ambipolar electric field for each $n_z$ in the optimization loop for a non-trace impurity, and the above approximation can be verified \emph{a posteriori} at each optimization step. We will however not do so in this work, but merely use the above argument to motivate why our optimized $n_z$ is not very sensitive to the specific values of the gradients in \eqref{eq:P}.

Note that while the optimized $n_z$ is insensitive to the gradients, the actual value of the peaking factor evaluated at this $n_z$ is sensitive to the gradients. 

\subsection{Optimization}
\label{sec:opt}
To avoid impurity accumulation, we seek to find an impurity density $n_z$ that minimizes \eqref{eq:P}. We restrict the problem to a finite number of degrees of freedom  by expressing $n_z$ in terms of a truncated Fourier-expansion
\begin{equation}
  \begin{aligned}
    n_z(\theta,\zeta) =& a_{00} f_{00}(\theta,\zeta) + \sum_{n=1}^{N} [a_{n0}f_{n0}(\theta,\zeta) +  b_{n0}g_{n0}(\theta,\zeta)] \\&+ \sum_{n=-N}^N \sum_{m=1}^{M} [a_{nm}f_{nm}(\theta,\zeta) +  b_{nm}g_{nm}(\theta,\zeta)],
  \end{aligned} \label{eq:fnz}
\end{equation}
where the basis functions
\begin{align}
  f_{nm}(\theta,\zeta) = 1 + \epsilon + \cos{(m\theta - N_\text{p}n\zeta)}, \\
  g_{nm}(\theta,\zeta) = 1 + \epsilon + \sin{(m\theta - N_\text{p}n\zeta)},
\end{align}
are chosen to be strictly positive ($\epsilon>0$, here $\epsilon=10^{-6}$), as the transport coefficients in \eqref{eq:DNz}--\eqref{eq:DPhi} diverge for $n_z=0$. Here, $\theta$ ($\zeta$) is the poloidal (toroidal) Boozer angle\cite{boozer1981,helanderStellaratorReview2014}, with $N_\text{p}$ the number of toroidal periods of the stellarator. Thus, we restrict ourselves to impurity densities that have the same discrete rotational symmetry as the magnetic field.
To avoid unrealistically sharp variation in $n_z$ and to limit the dimensionality of the problem, we restrict ourselves to $N=M=3$, which corresponds to 49 Fourier coefficients to optimize.

We eliminate one of the degrees of freedom, the $m=n=0$ mode, by specifying $\langle n_z \rangle$ on the flux-surface. The state-vector of the problem thus consists of the 48 unconstrained Fourier-modes. 
We furthermore will enforce $n_z>n_z^{\text{floor}} \geq 0$, which imposes a non-linear constraint on the Fourier-coefficients. The value of $n_z^{\text{floor}}$ can be tuned to restrict $n_z$ to a realistic range of values; the effect of changing $n_z^{\text{floor}}$ is investigated in appendix \ref{app:d}.

The optimization proceeds from an initial state-vector from which we find a \emph{local optimum} by applying the gradient-based \emph{method-of-moving-asymptotes}\cite{svanberg1987}, as implemented in the python version of the non-linear optimization package \textsc{NLopt}\cite{nlopt}. 

\section{Results}
\label{sec:res}

\begin{figure}
  \includegraphics[width=0.9\linewidth]{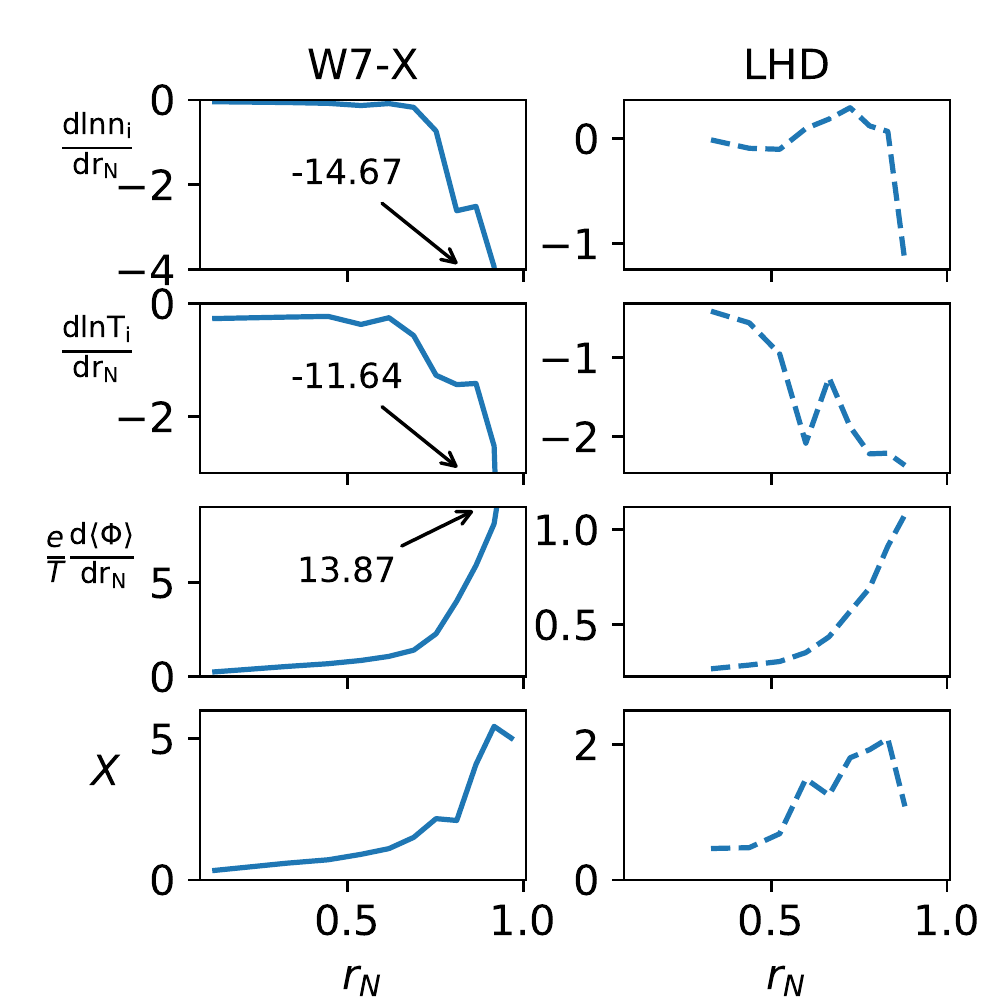}
  \put(-180,300){\Large a}
  \put(-180,230){\Large b}
  \put(-180,160){\Large c}
  \put(-180,80){\Large d}
  \put(-35,300){\Large e}
  \put(-35,230){\Large f}
  \put(-35,160){\Large g}
  \put(-35,90){\Large h}
  \caption{\label{fig:profiles} Profile gradients of the W7-X (solid line) and LHD (dashed) case under consideration. The arrows indicate the values at the outermost radial point. The quanitity $X$ is defined in \eqref{eq:X}. As $X$ is positive, minimizing the peaking factor should correspond to minimizing $D_{n_i}/D_{N_z}$.}
\end{figure}


\begin{figure*}
  \centering
  \includegraphics[width=0.75\linewidth]{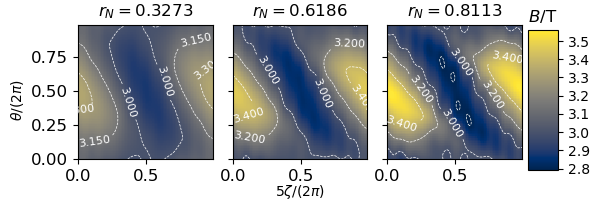}
  \put(-250,75){\Large \textcolor{White}{\bf a}}
  \put(-175,75){\Large \textcolor{White}{\bf b}}
  \put(-100,75){\Large \textcolor{White}{\bf c}}
  \caption{\label{fig:BW7X} Magnetic field strength $B$ for a W7-X standard configuration vacuum case over one field period.}

  \includegraphics[width=0.75\linewidth]{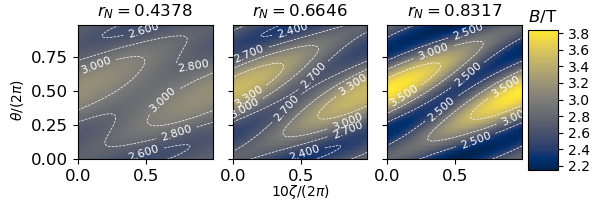}
  \put(-250,75){\Large \textcolor{White}{\bf a}}
  \put(-175,75){\Large \textcolor{White}{\bf b}}
  \put(-100,75){\Large \textcolor{White}{\bf c}}
  \caption{\label{fig:BLHD} Magnetic field strength $B$ for the LHD case\cite{velasco2017,mollen2018} over one field period.}
\end{figure*}

\begin{figure*}
  \includegraphics[width=0.75\linewidth]{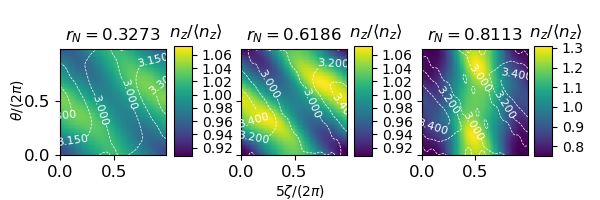}
  \put(-265,62){\Large \textcolor{White}{\bf a}}
  \put(-175,62){\Large \textcolor{White}{\bf b}}
  \put(-85,62){\Large \textcolor{White}{\bf c}}
\caption{\label{fig:nzW7X} $n_z$ optimized for minimum peaking factor, for the W7-X cases in \autoref{fig:BW7X}. The contours visualize the magnetic field at each flux-surface.}
  
\includegraphics[width=0.75\linewidth]{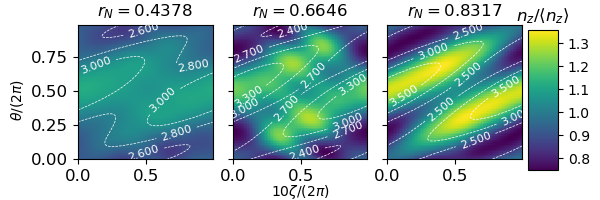}
\put(-250,75){\Large \textcolor{White}{\bf a}}
\put(-175,75){\Large \textcolor{White}{\bf b}}
\put(-100,75){\Large \textcolor{White}{\bf c}}
\caption{\label{fig:nzLHD} Figure corresponding to \autoref{fig:nzW7X}, but for the LHD case.}
\end{figure*}

We optimize $n_z$ for a W7-X and an LHD model case, with the logarithmic and potential gradients shown in \autoref{fig:profiles} and magnetic geometry in \autoref{fig:BW7X} and \autoref{fig:BLHD}. The gradients in the LHD case are taken from \nouncite{mollen2018}, while the W7-X gradients are from a simulated case in \nouncite{mollen2018}. The magnetic geometries are of a W7-X standard vacuum configuration\footnote{W7-X standard configuration available at (Verified 2019-02-15)\\\url{https://github.com/landreman/sfincs/blob/master/equilibria/w7x-sc1.bc}} and from LHD discharge \#113208 at $t=4.64\,\mathrm{s}$ \cite{velasco2017,mollen2018}.
Note that the cases considered here are not experimental cases, and merely use simulated or experimental data as realistic inputs.

From the last subfigure in \autoref{fig:profiles}, we see that $\left(\frac{e}{T} \frac{\d \langle\Phi\rangle}{\d r_\text{N}}  +  \frac{\d \ln n_i}{\d r_\text{N}} -0.5 \frac{\d \ln T}{\d r_\text{N}}\right) > 0$, which suggests that optimizing $n_z$ to minimize either the peaking factor $\mathcal{P}$ or $D_{n_i}/D_{N_z}$ are approximately equivalent.
The optimizations start from an initially homogeneous (i.e.\ flux-surface constant) $n_z$, and we restrict the flux-surface variations to $n_z>0.75 \langle n_z \rangle$, corresponding to $n_z^{\text{floor}}=0.75\langle n_z \rangle$.

The results of such an optimization for three different radii are shown in \autoref{fig:nzW7X} and \autoref{fig:nzLHD}, for the W7-X case and the LHD case, respectively. We see that the amplitude of the optimized $n_z$ appears to increase with radius.
For the W7-X case, the shape of the optimized $n_z$ are qualitatively different below and above $r_N \approx 0.62$. The corresponding potential variations are presented in appendix \ref{app:phi}. In both the LHD and W7-X for $r_N < 0.62$, the optimized $n_z$ are larger along the ridges of the maximum $B$, and peaking some distance away from the maximum value of $B$ -- sometimes with a second peak at the maximum of $B$.

\begin{figure}
  \includegraphics[width=0.95\linewidth]{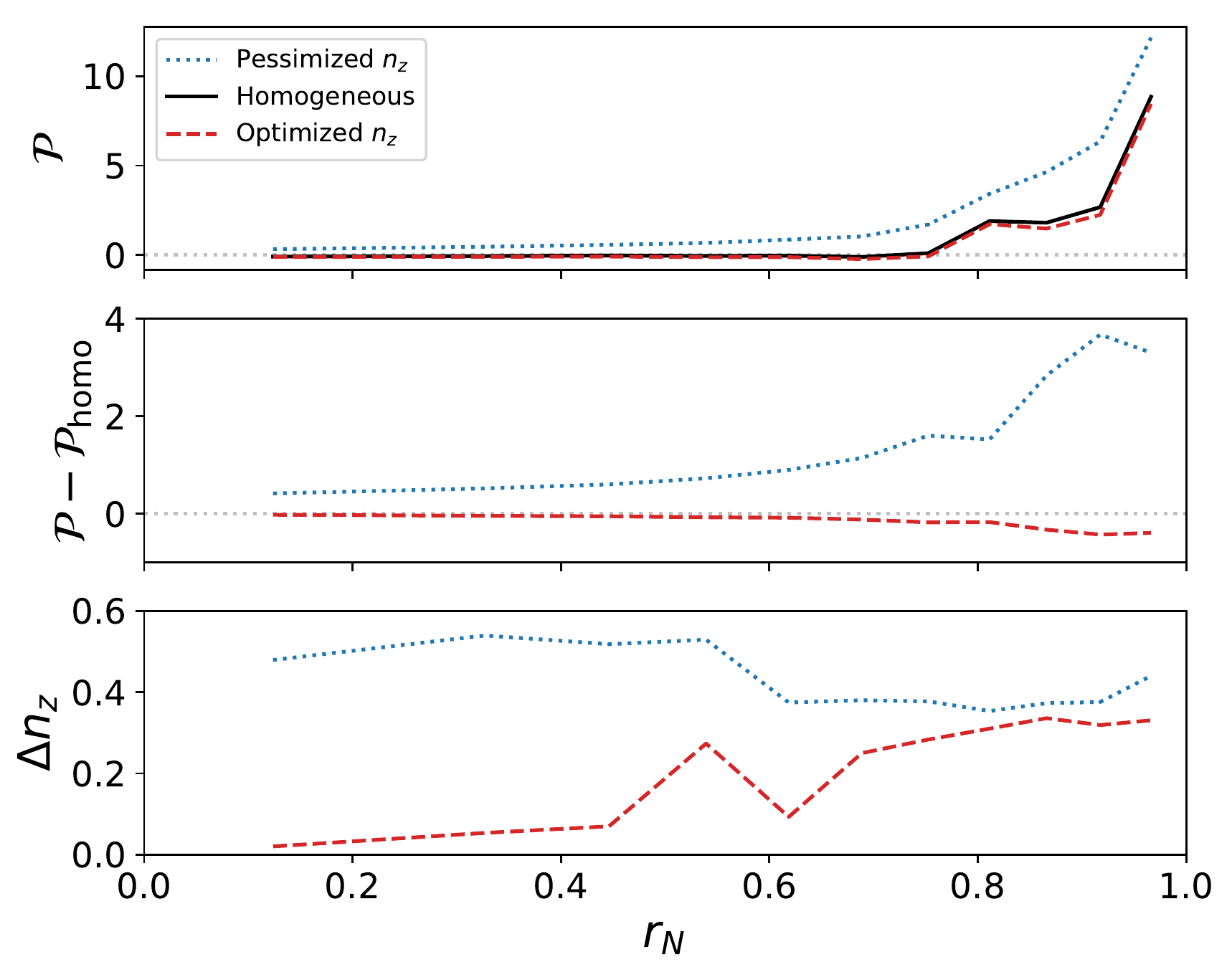}
  \put(-325,230){\Large a}
  \put(-325,190){\Large b}
  \put(-325,100){\Large c}
  \caption{\label{fig:peakW7X} \textbf{a:} peaking factor at different radii for the W7-X case.
    \textbf{b:} Differential change in peaking factor compared to the homogeneous case.
    \textbf{c:} The maximum deviation in $n_z/\langle n_z \rangle$.}

  \includegraphics[width=0.95\linewidth]{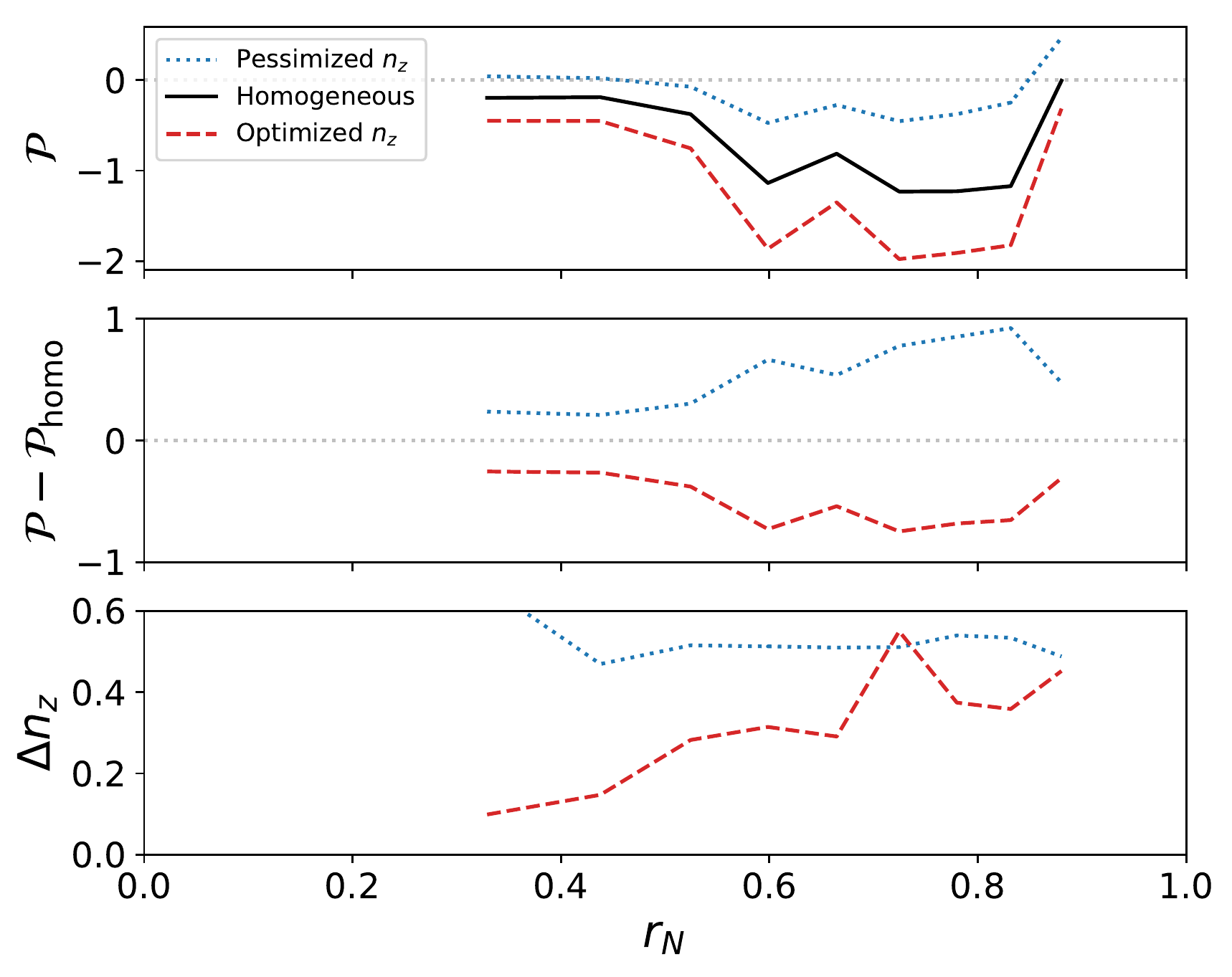}
  \put(-325,230){\Large a}
  \put(-325,190){\Large b}
  \put(-325,100){\Large c}
  \caption{\label{fig:peakLHD} Figure corresponding to \autoref{fig:peakW7X}, for the LHD case.}
\end{figure}

To evaluate the usefulness of targeting these optima, we need to know how much they reduce the peaking factor. This is shown in \autoref{fig:peakW7X}a and \autoref{fig:peakLHD}a, for the W7-X and LHD cases. As seen in \autoref{fig:peakW7X}a, the optimization only slightly reduces the peaking factor (red curve) compared to the homogeneous case (black) in the W7-X case. The difference between the optimized and homogeneous peaking factor, $|\mathcal{P}_{\text{opt}} - \mathcal{P}_{\text{homo}}|$, shown in \autoref{fig:peakW7X}b, is less than $0.1$ for $r_N<0.69$, and is dwarfed by the large peaking factor in the edge.
For the LHD case, the optimization yields a $|\mathcal{P}_{\text{opt}} - \mathcal{P}_{\text{homo}}|$ of at least $0.3$, and hence has a larger impact on the impurity accumulation in the core. This decrease is from an already negative $\mathcal{P}_{\text{homo}}$, which is due to difference in the geometry between the W7-X and LHD case.

In both cases, an unoptimized $n_z$ can result in a more peaked impurity profile, as seen from the pessimized curves in \autoref{fig:peakW7X}a and \autoref{fig:peakLHD}a -- which were obtained by maximizing, rather than minimizing, the peaking factor, where again we start from an initially homogeneous $n_z$. For the W7-X case, the effect of the pessimization is typically greater than the effect of optimization, with $\mathcal{P}_{\text{pes}} - \mathcal{P}_{\text{homo}}>0.3$ for all radius, while for the LHD case, $\mathcal{P} - \mathcal{P}_{\text{homo}}$ is comparable for optimization and pessimization, and ranges between $0.3$ to $1.0$.
This appears consistent with the findings in \nouncites{garcia2017,calvo2018nf,mollen2018}, where the transport of impurities tends to be more inwards when the effect of $n_z$ variation (through the electrostatic potential variation) is accounted for. It may also be consistent with \nouncite{yamaguchiIAEA2018}, where a beneficial effect of flux-surface variation is observed in simulations of the LHD. 

In \autoref{fig:peakW7X}c and \autoref{fig:peakLHD}c, we show the maximum deviations of $n_z$ from the flux-surface value
\begin{equation}
\Delta n_z = \max_{\zeta,\theta}\left(\left|\frac{n_z(\zeta,\theta)}{\langle n_z \rangle} - 1\right|\right),
\end{equation}
which illustrates that the amplitude of $n_z$ increases with radius, and that the nature of the optimized $n_z$ changes above $r_N \approx 0.62$ in the W7-X case, when a solution is found with much larger amplitude ($\Delta n_z \approx 0.30$, compared to $\Delta n_z = 0.09$ for $r_N\approx 0.62$). A similarly large amplitude also occurs at a lower, isolated radius ($r_N=0.52$) with no noticeable change in the peaking factor, indicating that the two kinds of optima have very similar peaking factors around these radii.
This shows that at least $\Delta n_z \sim 0.3$ variations can exist without severely impacting the peaking factor. On the other hand, the pessimized $n_z$ tend to have larger $\Delta n_z$ in both the W7-X and LHD case. 

\subsection{Influence of small-amplitude modes}
For the optimized $n_z$ to be reasonable targets in an experiment, they must be robust to small changes. We investigate this by considering how many of the 48 unconstrained Fourier modes actually contribute to lowering the peaking factor.

\begin{figure}
  \includegraphics[width=0.95\linewidth]{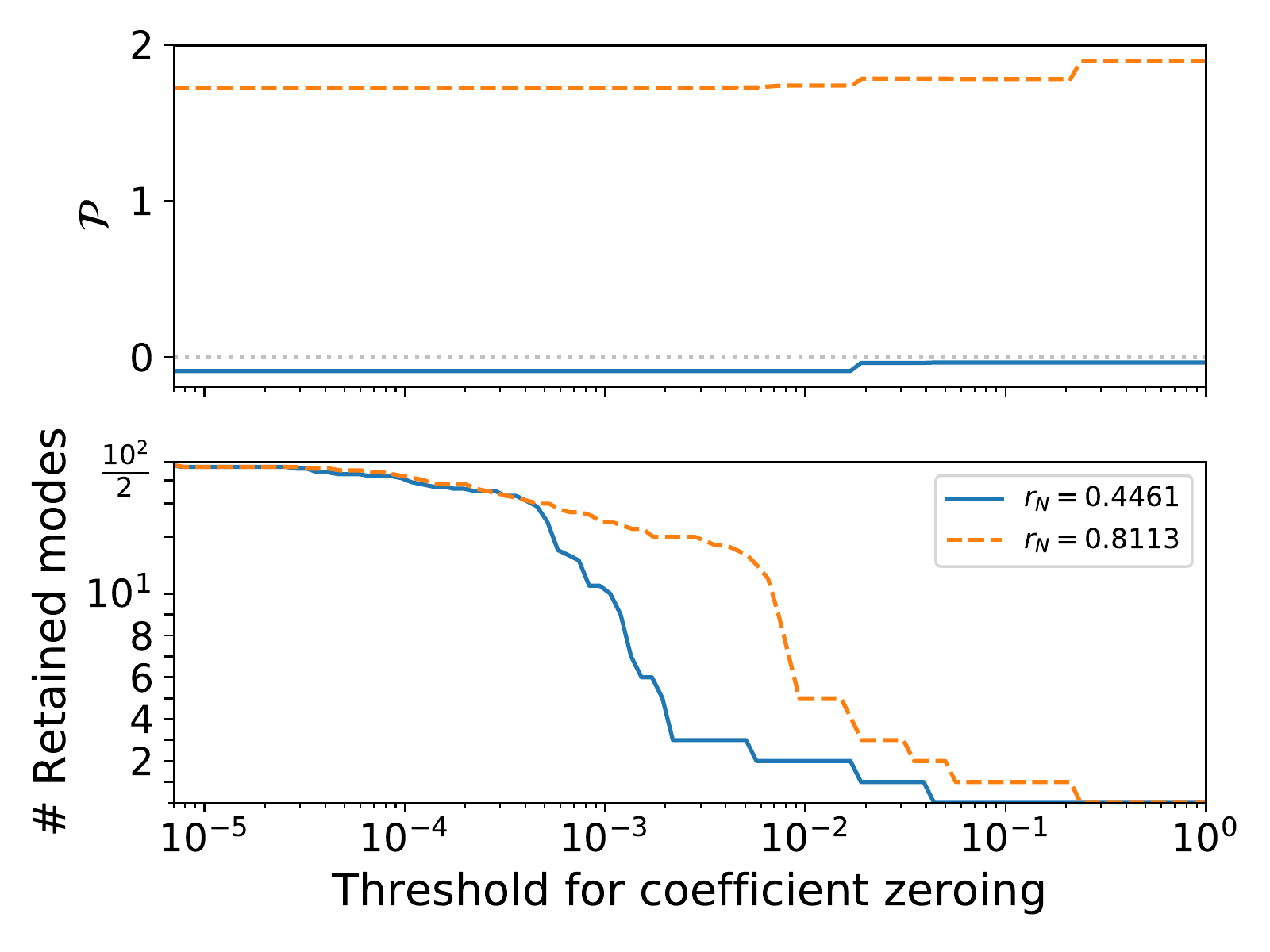}
  \put(-320,180){\Large a}
  \put(-320,130){\Large b}
  \caption{\label{fig:W7Xrs} The effect of zeroing out coefficients below a threshold value on the peaking factor $\mathcal{P}$ in the W7-X case.}

  \includegraphics[width=0.95\linewidth]{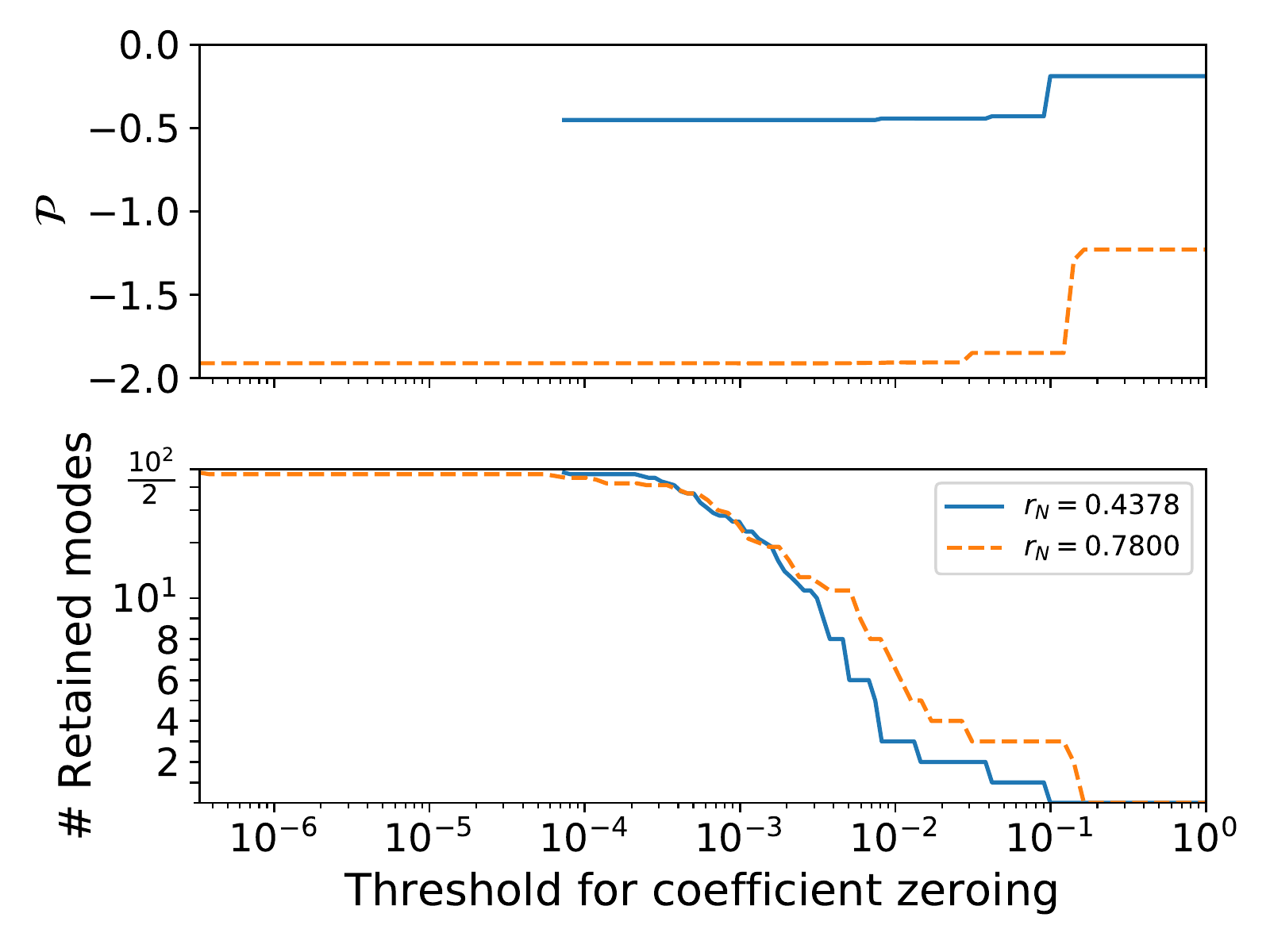}
  \put(-310,180){\Large a}
  \put(-310,130){\Large b}
  \caption{\label{fig:LHDrs} Figure corresponding to \autoref{fig:W7Xrs} for the LHD case.}
\end{figure}

We investigate this by zeroing out Fourier-coefficients below a threshold value in our representation of $n_z$, \eqref{eq:fnz}. The resulting change in peaking factor and the number of retained unconstrained modes is shown in \autoref{fig:W7Xrs} and \autoref{fig:LHDrs}, for the W7-X and LHD case at two different radii, which are representative of the behaviour in the core and the edge. Note that the constrained $m=0,n=0$ cosine mode is always retained, and is set by the other modes to ensure that $\langle n_z \rangle$ is held constant.
Also note the linear scale in the number of retained modes below 10 in \autoref{fig:W7Xrs} and \autoref{fig:LHDrs}, and the logarithmic scale when more than 10 modes are retained. For all cases, less than 10 modes actually contribute to visibly reducing the peaking factor, as seen by the constant peaking factor for low values of the zeroing-threshold. In fact,
the sharpest increase in the peaking factor occurs only when the last one or two unconstrained modes are zeroed out.
For the W7-X case, this sharp increase occurs when the last two (last) unconstrained mode is zeroed in the core (edge), while in the LHD case it occurs at the last (last two) in the core (edge).
Retaining only the two largest modes, $92\%$ ($90\%$) of the optimization in $\mathcal{P}_{\text{opt}} - \mathcal{P}_{\text{homo}}$ is retained in the LHD core (edge), while the corresponding number in the W7-X case is $99\%$ ($66\%$).
This shows that only one or two dominant unconstrained modes are needed for most of the optimization.

\begin{figure}
  \includegraphics[width=0.95\linewidth]{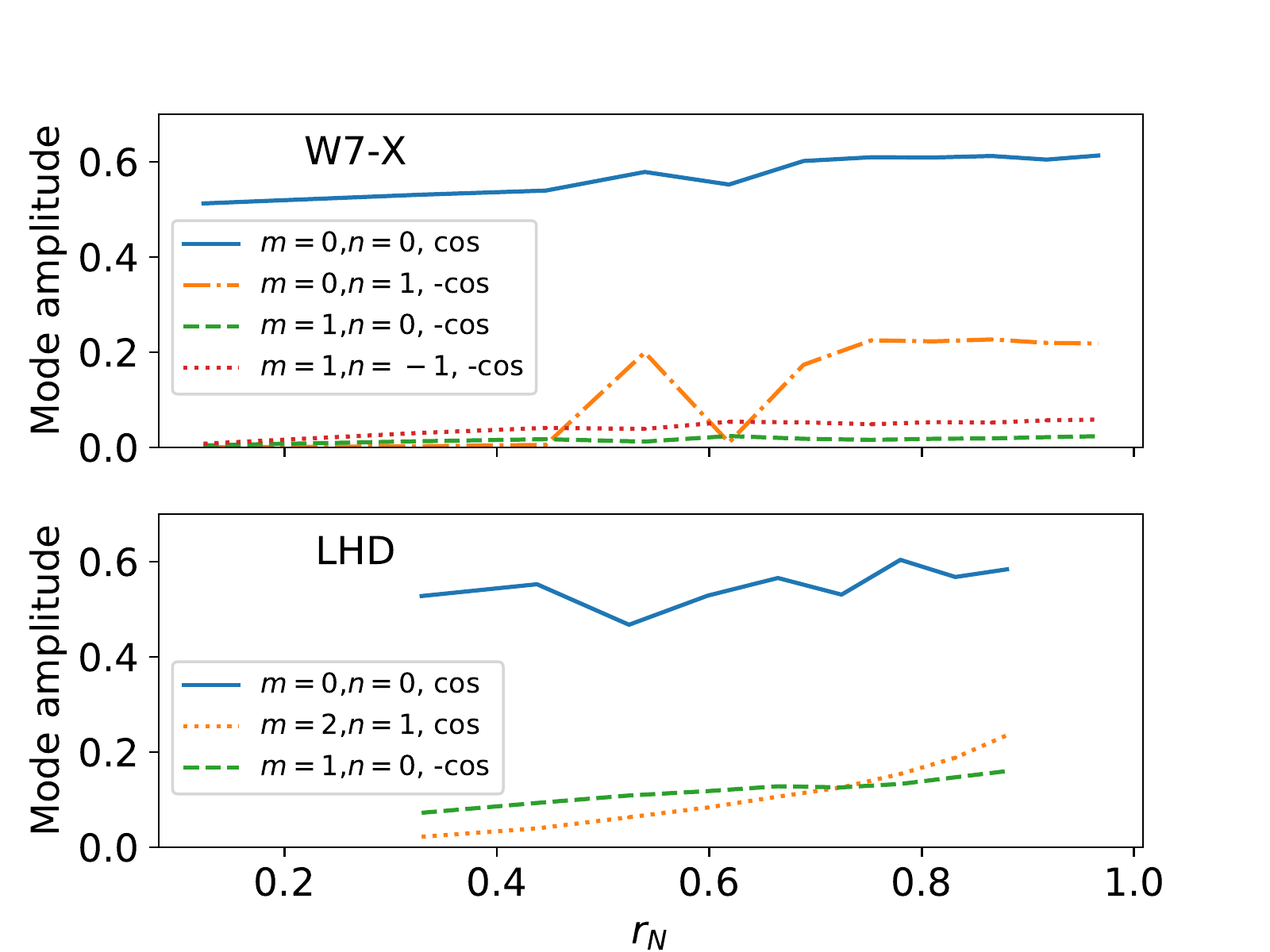}
  \put(-320,235){\Large a}
  \put(-320,110){\Large b}
  \caption{\label{fig:domModes} Dominant modes for the different radii in the W7-X and LHD cases. Modes with negative amplitudes are marked with a ``$-$'' in the legend.}
\end{figure}

To illustrate which two modes are the dominant ones, we plot the amplitude of the relevant modes against radius in \autoref{fig:domModes}, alongside the constrained $m=0,n=0$ mode. From the W7-X subfigure in \autoref{fig:domModes}, we clearly see that the nature of the optimized $n_z$ changes for $r_N > 0.62$, but it transitions temporarily already at $r_N \approx 0.54$, as previously observed. In the LHD case, the $m=1,n=0$ cosine mode is the dominant unconstrained mode at low $r_N$, but the $m=2,n=1$ cosine mode overtakes it at $r_N \approx 0.7$. In both cases, no sine modes contribute significantly to the optimization; sine modes appear to increase the peaking factor in the cases considered here.

By restricting our attention only to the two dominant modes, we can visualize the optimization landscape to identify the globally optimal amplitudes for these modes. For this purpose, we perform a two-dimensional scan in the amplitudes of the dominant modes for the W7-X and LHD cases. For the W7-X case, we restrict our attention to the two dominant modes at low $r_N$.
\begin{figure*}
  \includegraphics[width=0.75\linewidth]{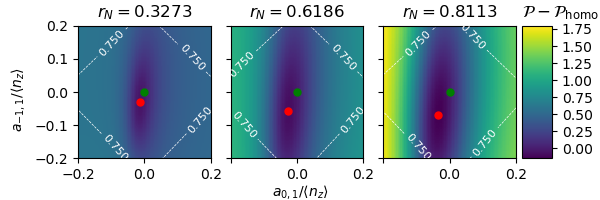}
  \put(-250,75){\Large \textcolor{White}{\bf a}}
  \put(-175,75){\Large \textcolor{White}{\bf b}}
  \put(-100,75){\Large \textcolor{White}{\bf c}}
  \caption{\label{fig:PW7X} The differential peaking factor $\mathcal{P}-\mathcal{P}_{\text{homo}}$ at each point in an amplitude scan of the two dominant modes at low $r_N$ in the W7-X optimization. Note that these two modes are not dominant in the previously found optima at $r_N>0.62$, see the W7-X subfigure of \autoref{fig:domModes}. The red dot indicates the location of the minimum, and the green dot a homogeneous $n_z$.}

  \includegraphics[width=0.75\linewidth]{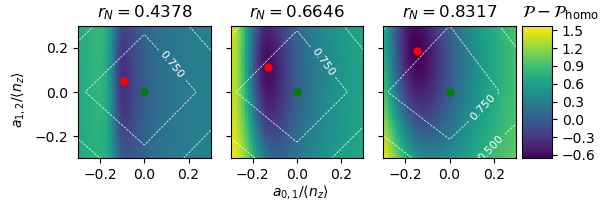}
  \put(-250,75){\Large \textcolor{White}{\bf a}}
  \put(-175,75){\Large \textcolor{White}{\bf b}}
  \put(-100,75){\Large \textcolor{White}{\bf c}}
  \caption{\label{fig:PLHD} A figure corresponding to \autoref{fig:PW7X}, for the LHD case.}
\end{figure*}

The peaking factor for each point in this scan is illustrated in \autoref{fig:PW7X} and \autoref{fig:PLHD}, for the W7-X and LHD case, respectively. The contour lines represent the limit imposed by the optimization requiring $n_z/\langle n_z \rangle>0.75$. From these figures, we see that the optimum (red dot) tends to higher amplitudes for higher $r_N$, and also that it is the local optimum starting from an initially homogeneous $n_z$ (green dot). The root mean square (RMS) of the amplitudes $\left(\sqrt{a_x^2 + a_y^2}/\langle n_z \rangle\right)$ scale linearly with $r_N$: for the W7-X case, the RMS amplitude at the different radii are $0.033$, $0.061$, and $0.079$, respectively, while they are $0.107$, $0.172$, and $0.237$ for the LHD case. In both the W7-X and LHD case, the optimum is relatively narrow in the amplitude of the $m=1,n=0$ cosine mode (x-axis), but broadens for higher $r_N$.

Regardless of how narrow the optimum is, it is always beneficial to have an $m=1,n=0$ cosine mode of moderate amplitude, compared to the homogeneous case, but the peaking factor rapidly increases when the amplitude becomes larger than a threshold value. For the W7-X case, the threshold values at these radii are $-0.02$, $-0.038$ and $-0.055$; and for the LHD case, they are $-0.13$, $-0.19$ and $-0.22$.  
The peaking factor is less sensitive to the amplitude of the other mode ($m=1,n=-1$ in W7-X; $m=2,n=1$ in LHD), but the minima in these broader basins nevertheless tend towards larger amplitudes at higher $r_N$.
For the LHD case, the amplitude of the minima becomes sufficiently large to hit the $n_z>0.75 \langle n_z \rangle$ constraint, which we imposed to exclude large $n_z$ variation. 

\subsection{Reduction in core density}
\label{sec:ricd}

The peaking factor can be integrated to yield the impurity pseudo-density profile
\begin{equation}
\left(\frac{N_z(r_N)}{N_z(r^\text{edge}_N)}\right)^{1/Z} = \exp{\left(\int\limits_{r_N}^{r^\text{edge}_N} \!\mathcal{P}(r_N')\, \d r_N'\right)},
\end{equation}
which allows us to assess how effective the optimization is in reducing the core impurity density. To perform the above integral numerically, we optimized the peaking factor for 97 radii in the W7-X case, and 47 radii in the LHD case, using the same modes as in \autoref{fig:PW7X} and \autoref{fig:PLHD} for the respective cases.

\begin{figure}
  \includegraphics[width=0.95\linewidth]{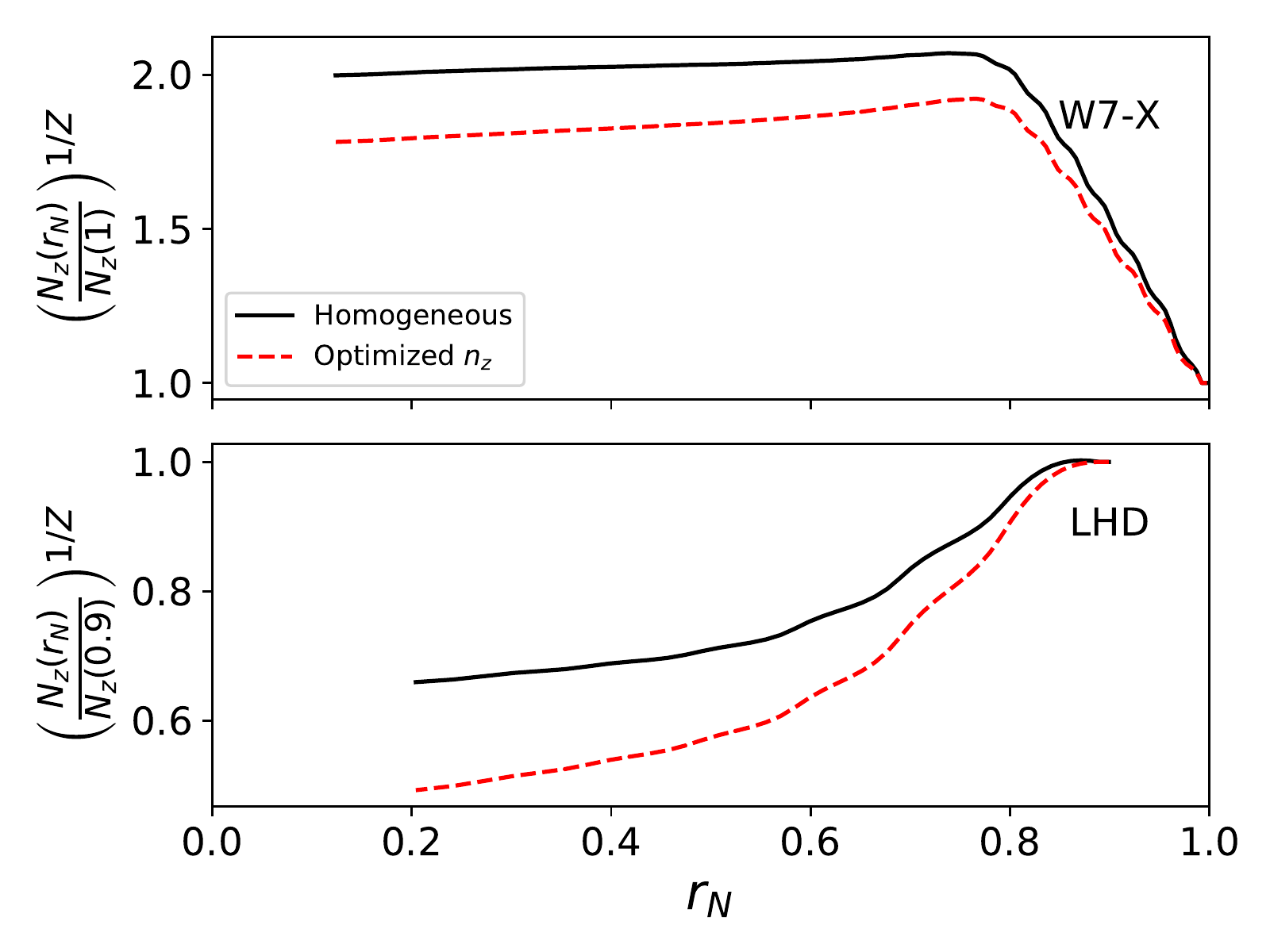}
   \put(-305,255){\Large a}
  \put(-305,130){\Large b}
  \caption{\label{fig:Nz} The steady-state impurity psuedo-density profiles $N_Z(r_N)$ for $n_z$ constant on the flux-surface (homogeneous) or $n_z$ optimized to minimize the peaking factor, for the W7-X case (a) and LHD case (b). The optimizations used the same modes as in \autoref{fig:PW7X} and \autoref{fig:PLHD} for the respective cases.}
\end{figure}

The resulting $N_z$ profiles are shown in \autoref{fig:Nz}. Thus, the optimization reduces $(N_z(r_N^{\text{core}})/N_z(r^{\text{edge}}_N))^{1/Z}$ to $89\%$ of the homogeneous $n_z$ value, from $2.00$ to $1.78$, in the W7-X case; and to $75\%$, from $0.659$ to $0.492$, in the LHD case. The corresponding reduction in the pseudo-density is exponentiated by $Z$: At $Z=24$, the optimization reduces the W7-X core impurity content to $6\%$ of the unoptimized levels, while for the LHD case, the corresponding reduction is to $0.09\%$  of the unoptimized levels. This shows that even modest reductions in $\mathcal{P}$ can yield large reductions in the impurity density at sufficiently large $Z$.

\section{Summary \& Conclusions}
We have performed numerical optimizations of the impurity density variation on the flux-surface, $n_z(\zeta,\theta)$, to minimize a semi-analytical expression for the impurity pseudo-density peaking factor, valid in the \emph{mixed-collisionality regime} with a collisional impurity of charge $Z\gg 1$ and a low-collisionality bulk-ion species. In order to constrain $n_z$ to a realistic range of variations, we have only performed local optimizations of $n_z$ around a homogeneous initial value, using a gradient-based minimization method to minimize the peaking factor. To further constrain $n_z$, we have imposed $n_z>0.75 \langle n_z \rangle$, as the optimization otherwise can produce unrealistically large $n_z$ variations on the flux-surface.

The results show that there is a potential to lower the $Z$-normalized impurity peaking factor by $0.3$ by controlling $n_z$ in the LHD case considered here, while the optimization in the W7-X case lowers the $Z$-normalized peaking factor by $0.1$ compared to a homogeneous $n_z$. On the other hand, we find that there are flux surface variations that lead to a $\sim 2.0$ increase in the $Z$-normalized impurity peaking factor in both the W7-X and LHD case. In both cases, the reduction in peaking factor can lead to significant reduction in the core density for sufficiently high impurity charge, with the ratio of optimized to homogeneous core densities being given by $N_z^{\text{opt}}/N_z^{\text{homo}} = 0.89^Z$ in the W7-X case and $0.75^Z$ in the LHD case considered here.

In conclusion, this study shows that it may be worthwhile to experiment with $n_z$ variation in stellarators to reduce core impurity accumulation.
The fact that only one or two modes of low mode-numbers are required to optimize the impurity density is promising, and could provide signatures of optimized impurity variations that may be targetable (by for example heating\cite{angioni2014,yamaguchiIAEA2018} or tailoring the magnetic field\cite{calvo2018b}) and measurable in experiments, using for example Doppler reflectometry\cite{estradaIAEA2018} or probes\cite{pedrosa2015}.
In particular, it appears that $m=1,n=0$ cosine modes of moderate amplitudes (up to $-0.07 \langle n_z \rangle$ to $-0.22 \langle n_z \rangle$, with larger amplitudes at larger radius) are beneficial in the LHD case considered here. For the W7-X standard configuration, the $n_z$ optimization is less effective in reducing the peaking factor, and from this perspective, it may be wise to avoid $n_z$ variations. On the other hand, there exist $n_z$ variations with large amplitudes and slightly-better than homogeneous performance in W7-X, so $n_z$ variations are not intrinsically harmful. Furthermore, at high enough $Z$, even a moderate reduction in peaking factor leads to a significant reduction in core density.
It is however unclear how these results extrapolate to different magnetic configurations, as this study is of limited scope. Furthermore, the $D_{N_z}$ coefficient may be modified by the presence of turbulent transport\cite{helander2018}, which should affect the results. These are interesting avenues for future work.


\section*{Acknowledgement}
This work has been carried out within the framework of the EUROfusion
Consortium and has received funding from the Euratom research and training
programme 2014-2018 and 2019-2020 under grant agreement No 633053. The views and
opinions expressed herein do not necessarily reflect those of the European
Commission.
SB and IP were supported by the International Career Grant of Vetenskapsr{\aa}det (Dnr.~330-2014-6313) and IP by Marie Sklodowska Curie Actions, Cofund, Project INCA 600398. SB's trip to the 27th IAEA Fusion Energy Conference was supported by The Royal Society of Arts and Sciences in Gothenburg. 
The authors would like to thank the LHD experiment group and the technical staff of LHD for their support of this work, and M.~Nunami for help with obtaining LHD data clearance.

\appendix
\section{Effects of changing the minimum $n_z$ value}
\label{app:d}

\begin{figure*}
\centering
\includegraphics[width=0.75\linewidth]{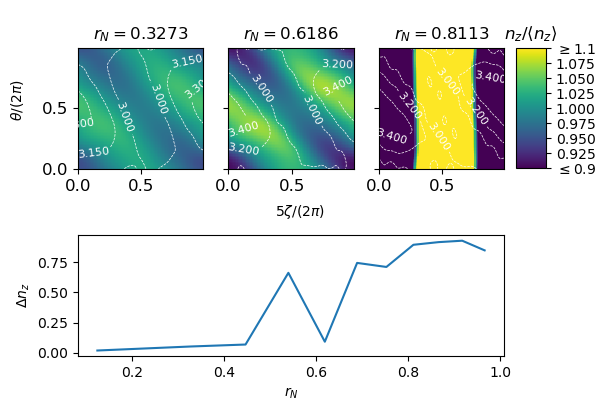}
\put(-250,160){\Large \textcolor{White}{\bf a}}
\put(-175,160){\Large \textcolor{White}{\bf b}}
\put(-100,160){\Large \textcolor{White}{\bf c}}
 \put(-250,70){\Large d}
  \caption{\label{fig:nzW7Xd0} \textbf{a-c:} $n_z$ optimized for minimum peaking factor with $n_z>0$, for the W7-X cases in \autoref{fig:BW7X}. The contours visualize the magnetic field at each flux-surface. \textbf{d:} $\Delta n_z$ for each flux-surface.}

  \includegraphics[width=0.75\linewidth]{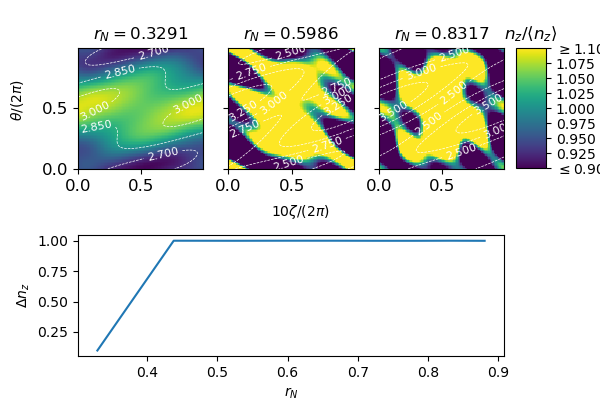}
  \put(-250,160){\Large \textcolor{White}{\bf a}}
\put(-175,160){\Large \textcolor{White}{\bf b}}
\put(-100,160){\Large \textcolor{White}{\bf c}}
 \put(-250,70){\Large d}
\caption{\label{fig:nzLHDd0} Figure corresponding to \autoref{fig:nzW7Xd0}, but for the LHD case.}
\end{figure*}

In the bulk of this paper, we optimized $n_z$ given the constraint $n_z>0.75 \langle n_z \rangle$. Here, we show the effects of lifting that constraint, to demonstrate why a constraint of this sort is necessary. 

In \autoref{fig:nzW7Xd0}a and \autoref{fig:nzLHDd0}a, we show the optimized $n_z$ obtained for the flux-surfaces in \autoref{fig:BW7X} and \autoref{fig:BLHD}, by the optimization procedure described in \autoref{sec:opt} starting from an initially homogeneous $n_z$ and only requiring $n_z>0$ everywhere on the flux-surface.

From these figures, we see that when the amplitude of the optimized $n_z$ is allowed to increase, the optimization algorithm starts to find a new kind of local minimum, with unrealistically large amplitude, a radically different shape, and no clear correlation to the magnetic field strength.
This occurs at $r_N > 0.62$ in the W7-X case, and already at $r_N > 0.33$ in the LHD case. 

\begin{figure}
  \includegraphics[width=0.75\linewidth]{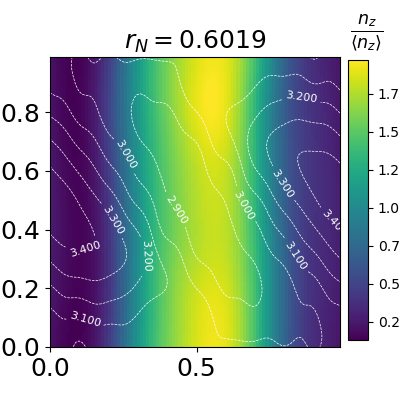}
  \caption{\label{fig:stochoptW7X} Optimized $n_z$ resulting from stochastic optimization in the W7-X case at $r_N=0.6$.}
\end{figure}

\begin{figure}
  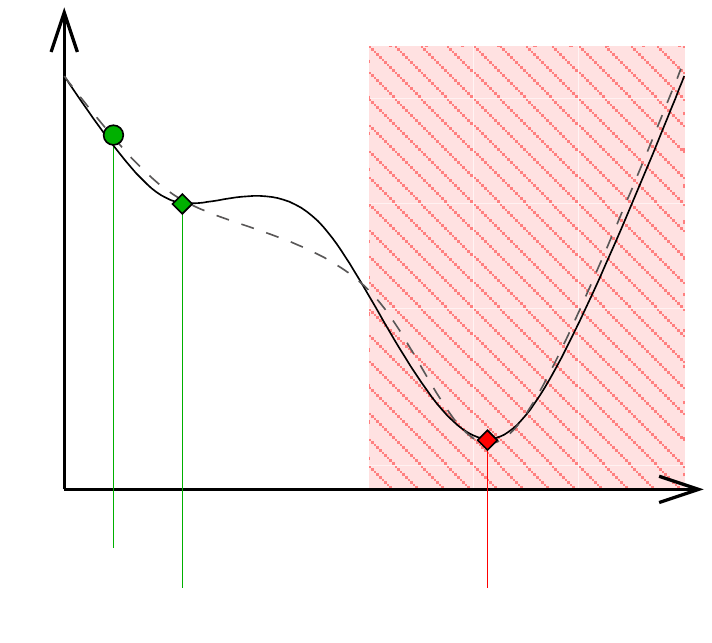
  \caption{\label{fig:stoill} Illustration of the local optimization landscape in the W7-X case, for two different radii. At higher $r_N$, a local optimizer starting at a uniform $n_z$ finds an optimum that was inaccessible but existed at lower $r_N$; as revealed by the stochastic optimization. The differences in peaking factor are here exaggerated for illustrative purposes.}
\end{figure}

Thus, without specifying a lowest allowed value for $n_z$, even optimizations starting from a homogeneous initial $n_z$ cannot be guaranteed to display reasonable $n_z$ variations.
The shape of the optima found at the outer radii in \autoref{fig:nzW7Xd0}a is in fact very similar to a ``global'' optimum obtained by stochastic optimization where several initial $n_z$ are randomly generated and the lowest local optimum resulting from performing local optimization from these initial $n_z$'s is taken as an upper bound for the global minimum. The result of this process for $r_N=0.6$ in W7-X is shown in \autoref{fig:stochoptW7X}. The situation in the W7-X case may thus be illustrated as in \autoref{fig:stoill}, where there always exist local minima unrealistically far from a homogeneous $n_z$ that are more optimal than those obtained from starting the local optimization at a homogeneous $n_z$.

This is the reason for optimizing on a more restrictive function-space, for example by specifying $n_z>n_z^{\text{floor}}$ for some $n_z^{\text{floor}}$. A better approach would be to restrict $n_z$ to a function-space of ``reasonable'' impurity density variations based on the physical modeling of the phenomena causing the flux-surface variation, but this is outside the scope of this work.

\section{Corresponding potential variation}
\label{app:phi}
\begin{figure*}
  \includegraphics[width=0.75\linewidth]{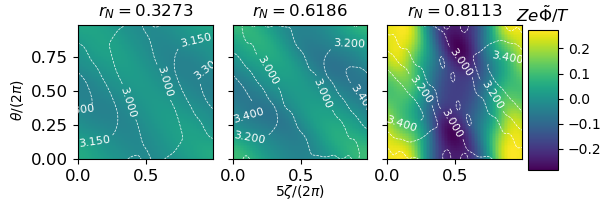}
  \put(-250,75){\Large \textcolor{White}{\bf a}}
  \put(-175,75){\Large \textcolor{White}{\bf b}}
  \put(-100,75){\Large \textcolor{White}{\bf c}}
  \caption{\label{fig:Phi1W7X} $\tilde{\Phi}$ corresponding to the W7-X $n_z$ in \autoref{fig:nzW7X}, which is optimized for minimum peaking factor, for the W7-X cases in \autoref{fig:BW7X}. The contours visualize the magnetic field at each flux-surface.}

  \includegraphics[width=0.75\linewidth]{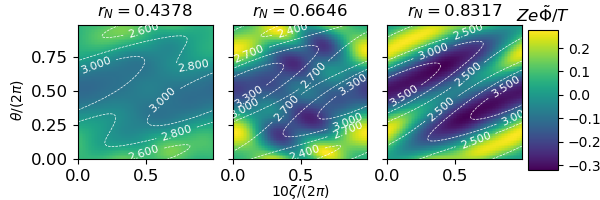}
  \put(-250,75){\Large \textcolor{White}{\bf a}}
  \put(-175,75){\Large \textcolor{White}{\bf b}}
  \put(-100,75){\Large \textcolor{White}{\bf c}}
  \caption{\label{fig:Phi1LHD} Figure corresponding to \autoref{fig:Phi1W7X} for the LHD case, showing the $\tilde{\Phi}$ due to the $n_z$ in \autoref{fig:nzLHD}.}
\end{figure*}

In this work, we calculate $n_z$ that optimize the impurity peaking factor, with little regard for the physical mechanisms that set $n_z$. However, physics-based calculations of $n_z$ often calculate flux-surface variation of the electrostatic potential, from which $n_z$ is obtained through \eqref{eq:boltzmann}. To connect our results to previous theoretical studies, we thus invert \eqref{eq:boltzmann} to obtain $\tilde{\Phi}$'s corresponding to our optimized $n_z$
\begin{equation}
\frac{Ze\tilde{\Phi}}{T} = -\log{n_z} + C,
\end{equation}
where $C$ is a constant calculated to make $\langle\tilde{\Phi} \rangle = 0$. The resulting $Ze\tilde{\Phi}/T$ for the W7-X and LHD cases are displayed in \autoref{fig:Phi1W7X} and \autoref{fig:Phi1LHD}.

\bibliography{plasma-bib.bib} 
\end{document}